\documentclass[showpacs,twocolumn,prl,bibnotes,10pt,letterpaper]{revtex4}%
\usepackage{amsfonts}
\usepackage{amsmath}
\usepackage{amssymb}
\usepackage{graphicx}%
\begin{document}
\title{Coupled quantum electrodynamics in photonic crystal nanocavities}
\author{Y.-F. Xiao$^{1,2}$}
\email{yfxiao@ustc.edu.cn}
\author{J. Gao$^{1}$}
\author{X.-B. Zou$^{2}$}
\author{J. F. McMillan$^{1}$}
\author{X. Yang$^{1}$}
\author{Y.-L. Chen$^{2}$}
\author{Z.-F. Han$^{2}$}
\author{G.-C. Guo$^{2}$}
\author{C. W. Wong$^{1}$}
\email{cww2104@columbia.edu}
\affiliation{$^{1}$Optical Nanostructures Laboratory, Columbia University, New York, NY 10027}
\affiliation{$^{2}$Key Laboratory of Quantum Information, University of Science and
Technology of China, P. R. China}


\begin{abstract}
We show that a scalable photonic crystal nanocavity array, in which single
embedded quantum dots are coherently interacting, can perform as a universal
single-operation quantum gate. In a passive system, the optical analogue of
electromagnetically-induced-transparency is observed. The presence of a single
two-level system in the array dramatically controls the spectral lineshapes.
When each cavity couples with a two-level system, our scheme achieves
two-qubit gate operations with high fidelity and low photon loss, even in the
bad cavity limit and with non-ideal detunings.

\end{abstract}
\maketitle

\textit{Introduction.}---Cavity quantum electrodynamics (QED) describes a
small number of atoms strongly coupling to quantized electromagnetic fields
through dipolar interactions inside an optical cavity. Up to now, it is one of
few experimentally realizable systems in which the intrinsic quantum
mechanical coupling dominates losses due to dissipation, providing an almost
ideal system which allows quantitative studying of a dynamical open quantum
system under continuous observation. In the strongly coupled regime, quantum
state mapping between atomic and optical states becomes possible, which has
promising implications towards realization of chip-scale quantum information
processing \cite{science 02}. Over the past few years, theoretical and
experimental interests are mainly focused on a single cavity interacting with
atoms, and important successes have been made ranging from trapping of
strongly coupled single atoms inside an optical microcavity \cite{McKeever}
and deterministic generation of single-photon states \cite{Keller}, to
observation of atom-photon quantum entanglement \cite{Volz} and implementation
of quantum communication protocols \cite{Rosenfeld}.

For more applications in cavity QED, current interest lies in the coherent
interaction among several distant cavities. The coherent interaction of cavity
arrays has been studied as a classical analogy to electromagnetically induced
transparency (EIT) in both theory \cite{Smith,Xiao01} and experiment
\cite{Xu,Totska}. Coupled cavities can be utilized for coherent optical
information storage because they provide almost lossless guiding and coupling
of light pulses at ultrasmall group velocities. When dopants such as atoms or
quantum dots (QDs) interacts with these cavities, the spatially separated
cavities have been proposed for implementing quantum logic and constructing
quantum networks \cite{Cirac}. Recent studies also show a photon-blockade
regime and Mott insulator state \cite{Hartmann}, where the two-dimensional
hybrid system undergoes a characteristic Mott insulator (excitations localized
on each site) to superfluid (excitations delocalized across the lattice)
quantum phase transition at zero temperature \cite{Greentree}. The character
of a coupled cavity configuration has also been studied using the photon Green
function \cite{Hughes,Hu}.

\begin{figure}[pb]
\centerline{\includegraphics[keepaspectratio=true,width=0.4\textwidth]{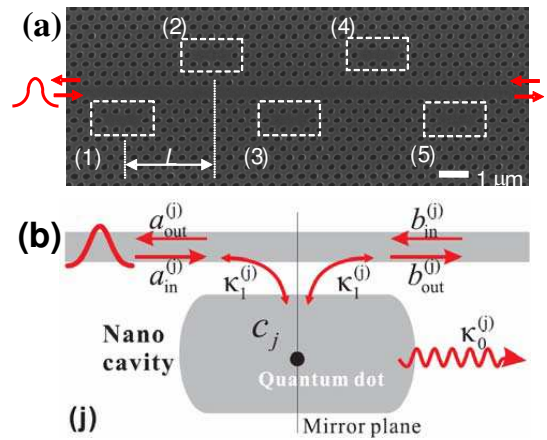}}\caption{(color)
(a) Example scanning electronic micrograph of periodic waveguide-resonator
structure containing $N$ side-coupled nanocavities (\textit{h}-polarized) at a
distance $L$. The nanocavities are side coupled through the integrated
waveguide, with no direct coupling between any two nanocavities. (b) The
\textit{j}-th QD-cavity subsystem.}%
\end{figure}

\textit{Theoretical model.}---In this paper, using transmission theory we
study coherent interactions in a cavity array which includes $N$ nanocavity-QD
subsystems, with indirect coupling between adjacent cavities through a
waveguide (Fig. 1), and examine its implementation as a two-qubit
single-operation quantum gate. Recent experimental efforts have reported
remarkable progress in solid-state nanocavities, such as an ultrahigh quality
factor \cite{Noda,Tanabe}, observation of strong coupling and vacuum Rabi
splitting \cite{Yoshie,Hennessy}, transfer of single photons between two
cavities \cite{Englund}, and deterministic positioning of a cavity mode with
respect to a QD \cite{Badolato}.

First, we investigate a subsystem in which a single nanocavity interacts with
an isolated QD. Here for simplicity we suppose that only a single optical
resonance mode (\textit{h}-polarized) is present in the nanocavity, although
two-mode cavity-QD interactions have been considered earlier \cite{Xiao02}.
The nanocavity-QD-waveguide subsystem has mirror-plane symmetry, so that the
mode is even with respect to the mirror plane. Omitting the terms which
concern the Langevin noises, we can easily obtain Heisenberg equations of
motion \cite{Waks}%
\begin{align}
\frac{{dc_{j}}}{{dt}}  &  =-\mathrm{{i}}\left[  {c_{j},H_{j}}\right]
-\Gamma_{j}c_{j}+\mathrm{{i}}\sqrt{\kappa_{1,j}}\left(  {a_{\mathrm{{in}}%
}^{\mathrm{{(j)}}}+b_{\mathrm{{in}}}^{\mathrm{{(j)}}}}\right)  ,\\
\frac{{d\sigma_{-,j}}}{{dt}}  &  =-\mathrm{{i}}\left[  {\sigma_{-,j},H_{j}%
}\right]  -\gamma_{j}\sigma_{-,j},
\end{align}
where $c_{j}$ is bosonic annihilation operator of the \textit{j}-th cavity
mode with resonant frequency $\omega_{c,j}$. $a_{\mathrm{{in}}}^{\mathrm{{(j)}%
}}(b_{\mathrm{{in}}}^{\mathrm{{(j)}}})$ and $a_{\mathrm{{out}}}^{\mathrm{{(j)}%
}}(b_{\mathrm{{out}}}^{\mathrm{{(j)}}})$ describe the input and output fields
in the left (right) port respectively. $2\Gamma_{j}$ represents total cavity
decay with $\Gamma_{j}=\left(  {\kappa_{0,j}+2\kappa_{1,j}}\right)  /2$, where
$\kappa_{0,j}$ is the intrinsic cavity decay rate and $\kappa_{1,j}$ the
external cavity decay rate. $\sigma_{-(+),j}$ is the descending (ascending)
operator of the interacting two-level QD with transition frequency
$\omega_{r,j}$. $\gamma_{j}$ is the total decay rate of the QD, including the
spontaneous decay (at rate $\gamma_{s}$) and dephasing (at rate $\gamma_{p}$)
in the excited state $\left\vert e\right\rangle $; $H_{j}$ is the subsystem
Hamiltonian $H_{j}=\omega_{c,j}c_{j}^{\dag}c_{j}+\omega_{r,j}\sigma_{+,j}%
c_{j}+\left[  {g_{j}\left(  {\vec{r}}\right)  \sigma_{+,j}c_{j}+h.c.}\right]
$, where ${g_{j}\left(  {\vec{r}}\right)  }$ is the coupling strength between
the cavity mode and the dipolar transition $\left\vert g\right\rangle
\leftrightarrow\left\vert e\right\rangle $.

Our scheme operates in the weak excitation limit (excitated by a weak
monochromatic field or a single photon pulse with frequency $\omega$), so that
the motion equations can be solved, with the transport relation%
\begin{equation}
\left(
{\begin{array}{*{20}c} {b_{{\rm{in}}}^{{\rm{(j)}}} \left( \omega \right)} \\ {b_{{\rm{out}}}^{{\rm{(j)}}} \left( \omega \right)} \\ \end{array}}%
\right)  =T_{j}\left(
{\begin{array}{*{20}c} {a_{{\rm{in}}}^{{\rm{(j)}}} \left( \omega \right)} \\ {a_{{\rm{out}}}^{{\rm{(j)}}} \left( \omega \right)} \\ \end{array}}%
\right)  .
\end{equation}
Here the transport matrix is
\begin{equation}
T_{j}=\frac{1}{{\alpha_{j}+\kappa_{1,j}-\Gamma_{j}}}\left(
{\begin{array}{*{20}c} { - \kappa _{1,j} } & {\alpha _j - \Gamma _j } \\ {\alpha _j - \Gamma _j + 2\kappa _{1,j} } & {\kappa _{1,j} } \\ \end{array}}%
\right)  ,
\end{equation}
where $\alpha_{j}=\mathrm{{i}}\Delta_{c,j}+\left\vert {g_{j}\left(  {\vec{r}%
}\right)  }\right\vert /\left(  {\mathrm{{i}}\Delta_{r,j}-\gamma_{j}}\right)
$, and $\Delta_{c,j}=\omega-\omega_{c,j}$ ($\Delta_{r,j}=\omega-\omega_{r,j}$)
represents the detuning between the input field and the cavity mode (QD
transition). The transport matrix can be regarded as a basic cell in cascading
the subsystems and obtaining the whole transportation for the $N$-coupled
cavity-QD system. The transport properties can thus be expressed as%
\begin{equation}
\left(
{\begin{array}{*{20}c} {b_{{\rm{in}}}^{{\rm{(N)}}} \left( \omega \right)} \\ {b_{{\rm{out}}}^{{\rm{(N)}}} \left( \omega \right)} \\ \end{array}}%
\right)  =T_{N}T_{0}\cdots T_{0}T_{2}T_{0}T_{1}\left(
{\begin{array}{*{20}c} {a_{{\rm{in}}}^{{\rm{(1)}}} \left( \omega \right)} \\ {a_{{\rm{out}}}^{{\rm{(1)}}} \left( \omega \right)} \\ \end{array}}%
\right)  ,
\end{equation}
where $T_{0}$ is the transport matrix via the waveguide with a propagation
phase $\theta$. Eq. (5) is an important result in this paper, and in the
following we will show that it can be used in various interesting physical processes.

\textit{Spectral character of coupled QD-cavity arrays.}---To examine the
physical essence, we first examine the spectral character of the coupled
QD-cavity system. The reflection and transmission coefficients are defined as
$r_{N1}\equiv a_{\mathrm{{out}}}^{(1)}/a_{\mathrm{{in}}}^{(1)}$ and
$t_{N1}\equiv b_{\mathrm{{out}}}^{(N)}/a_{\mathrm{{in}}}^{(1)}$. We also
assume that these cavities possess the same dissipation characteristic, i.e.,
$\kappa_{0,j}=\kappa_{0}$, $\kappa_{1,j}=\kappa_{1}$, $\kappa_{1}=50\kappa
_{0}$, and $\Gamma_{j}=\Gamma$.

\begin{figure}[tb]
\centerline{\includegraphics[keepaspectratio=true,width=0.4\textwidth]{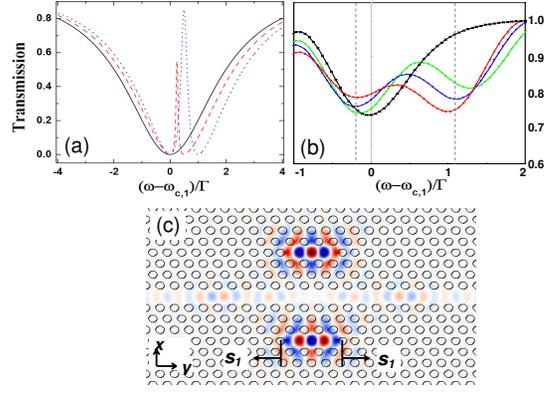}}\caption{(color)
(a) and (b): Transmission spectra of two coupled empty nanocavities. Solid,
dashed, and dotted lines describe the cases of $\delta_{21}=0,\Gamma/2,\Gamma
$, respectively. The used parameters: $\gamma=\kappa_{0},\theta=20\pi$. (b)
Numerical 3D FDTD simulations of optical analogue of EIT in two coherently
coupled cavity ($\theta=0$) for detunings $1.14\Gamma$ (red; $\Delta
\varepsilon_{\mathrm{cavities}}=0.135$), $1.26\Gamma$ (blue; $\Delta
\varepsilon_{\mathrm{cavities}}=0.160$), and $1.49\Gamma$ (green;
$\Delta\varepsilon_{\mathrm{cavities}}=0.185$). The arrows denote the EIT peak
transmissions. The dashed grey lines denote the two detuned individual
resonances for the case of $s_{1}=0.05a$. The black curve is for a single
cavity transmission for reference. (c) Example $E_{x}$-field distribution of
coupled empty photonic crystal nanocavities.}%
\end{figure}

Fig. 2a describes the transmission spectra of two coupled empty cavities with
different detuning ($\delta_{21}\equiv\omega_{c,2}-\omega_{c,1}$). When the
two cavities are exactly resonant, a transmission dip is observed; with
increasing$\ \delta_{21}$, a sharp peak exists at the center position between
the two cavity modes. This is directly analogous to the phenomenon of EIT in
atomic vapors, and examined exactly through ab initio 3D finite-difference
time-domain (FDTD) numerical simulations. Specifically, Fig. 2b shows an
example of the field distributions through the coherent interaction with two
coupled empty (without QD) cavities, where the resonance of one cavity is
detuned by three cases: $\delta_{21}=1.14\Gamma$, $1.26\Gamma$, and
$1.49\Gamma$. The optical EIT-like resonance is observed on top of a
background Fabry-Perot oscillation (due to finite reflections at the waveguide
facets). The analogy and difference between optical and atomic EIT are
recently discussed in Ref. \cite{Xiao01}.

In the presence of QDs, Fig. 3a (top) shows the spectral characteristics in
which a single QD resonantly interacts with the first cavity. When both
cavities are resonant, there exist two obvious sharp peaks located
symmetrically around $\omega=0$ (For convenience, we define $\omega_{c,1}=0$).
This fact can be explained by the dressed mode theory. Resonant QD-cavity
interaction results in two dressed cavity modes, which are significantly
detuned from the second cavity mode with the detuning $\pm\left\vert
{g_{1}(\vec{r})}\right\vert =\pm\Gamma/2$. Both dressed modes non-resonantly
couple with the second cavity mode, resulting in two EIT-like peaks located at
frequencies $\omega\approx\pm\Gamma/4$. When $\delta_{21}=\Gamma/2$, one
dressed cavity mode non-resonantly couples with the second cavity mode with a
detuning $\Gamma$, which leads to an EIT-like peak located near $\omega
\simeq0$; while the other dressed mode resonantly couples with the second
cavity mode, which does not result in an EIT-like spectrum. When $\delta_{21}$
continually increases, e.g., $\delta_{21}=\Gamma$, the vanished peak reappears
since the two dressed modes are always non-resonant with the second cavity
mode. Fig. 3a (bottom) illustrates the case where both cavities resonantly
interact with a single QD each. Similar to the above analysis, we can explain
the number and locations of sharp peaks with respect to different $\delta
_{21}$ by comparing the two pairs of dressed cavity modes. For example, when
$\delta_{21}=\Gamma$, the dressed modes in the first cavity is located at
$\pm\Gamma/2$ while the second pair is at $\Gamma/2$ and $3\Gamma/2$, so that
the EIT-like peaks must locate $[-\Gamma/2,\Gamma/2]$ and $[\Gamma
/2,3\Gamma/2]$, i.e., two peaks are near $0$ and $\Gamma$. Fig. 3b shows the
spectral character of three coupled QD-cavity subsystems, under various
cavity-cavity and QD-cavity detunings that might be observed experimentally,
involving Autler-Townes splittings and Fano interferences.

To further examine this coupled QD-cavity system, Fig. 3c shows the
transmission phase shift for various qubit detunings, where the cavity and QD
transition are resonant for both subsystems. We find that the phase shift has
a steep change, which corresponds to a strong suppression of the flying-qubit
group velocity. As shown in Fig. 3d, the delay time ($\tau_{\mathrm{sto}}$) in
this coupled system is almost hundreds of the cavity lifetime ($\tau
_{\mathrm{life}}=1/2\Gamma$). This proposed coupled QD-cavity system can
essentially be applied in storing quantum information of light. Although
having shorter coherence times than atomic ion qubit memory \cite{Langer}, our
solid-state implementation has an achievable bandwidth of $\sim50$ \textrm{MHz
}in contrast to less than $100$ \textrm{kHz} in atomic systems, with a
comparable delay-bandwidth product. One can also consider dynamical tuning
\cite{Yanik} to tune the cavity resonances with respect to the QD transitions
to break the delay-bandwidth product in our solid-state cavity array system.

\begin{figure}[tb]
\centerline{\includegraphics[keepaspectratio=true,width=0.45\textwidth]{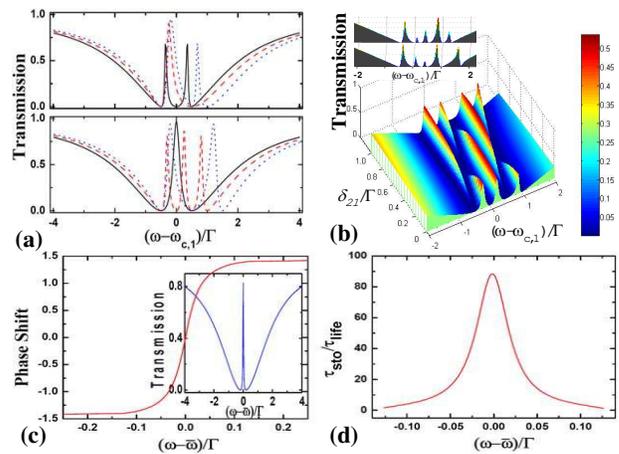}}\caption{(color)
(a) Transmission spectra of two coupled subsystems with one QD (top) and two
QDs (bottom) where $g=\Gamma/2$. Other conditions are same as Fig. 2a. (b)
Spectral character of three coupled subsystems with $\delta_{31}=\Gamma/2$.
Inset: $\delta_{31}=\Gamma/2$, $\delta_{21}=0$, $\delta_{1}=\delta_{3}=0$,
with $\delta_{2}=\Gamma/2$ (top) and $\Gamma$ (bottom), where $\delta
_{j}\equiv\omega_{c,j}-\omega_{r,j}$. (c) and (d): Photon phase shift and
delay ($\tau_{\mathrm{sto}}$) through two QD-nanocavity subsystems, where
$\omega_{c(r),j}=\varpi$, $g=0.2\Gamma$. Inset: transmission spectrum.}%
\end{figure}

\textit{Quantum phase gate operation.}---Now we show the possibility of
quantum gate operation based on the dynamical evolution. In constructing
quantum logic system, we have stationary qubits represented by two ground
states $\left\vert g\right\rangle $ and $\left\vert r\right\rangle $ of QDs.
The two ground states can be obtained via QD spin-states such as shown
remarkably in Ref. \cite{Atatuer} with near-unity fidelity. To facilitate the
discussion but without loss of generality, we consider an all resonance case
(i.e., $\omega=\omega_{c(r),j}$) to describe idea of the phase gate operation,
and in the subsequently numerical calculation, we will demonstrate the gate
feasibility under non-ideal detunings.

As an example, we focus on how to unconditionally realize a stationary
two-qubit (two QDs) phase gate. The input weak photon pulse is assumed
\textit{h}-polarized and we should discuss the following cases with different
initial QD states. Case I: The two QDs are initially prepared in $\left\vert
u\right\rangle _{1}\left\vert v\right\rangle _{2}(u,v=g,r)$ and at least one
QD occupies the ground state $\left\vert r\right\rangle $. It is not difficult
to find $r_{21}\simeq-1$ and $t_{21}\simeq0$ under the over-coupling regime
($\kappa_{0}\ll\kappa_{1}$) and with large Purcell factor ($g^{2}/\Gamma
\gamma\gg1$). This fact can be understood by regarding the resonant condition
($\omega_{c,j}=\omega$). The input photon will be almost reflected by one
\textit{empty} cavity, in which the QD is in $\left\vert r\right\rangle $,
resulting in a final state $-\left\vert u\right\rangle _{1}\left\vert
v\right\rangle _{2}\left\vert R\right\rangle $, where $\left\vert
R\right\rangle $ denotes the reflected photon. Case II: The QDs is initially
prepared in $\left\vert g\right\rangle _{1}\left\vert g\right\rangle _{2}$.
With the over-coupling lifetime and the large Purcell factor, we have
$r_{21}\simeq0$ and $t_{21}\simeq1$, so that the photon pulse passes through
the two cavities in turn via the waveguide, and the resulting state is
$\left\vert g\right\rangle _{1}\left\vert g\right\rangle _{2}\left\vert
T\right\rangle $, where $\left\vert T\right\rangle $ describes the transmitted
photon. Here note that the spatial mode of the output photon is actually
entangled with the QD states. To construct the final gate operation, it is
necessary to remove the distinguishability of the two photon spatial modes
(transmitted and reflected). Here we introduce a reflecting element in the end
of waveguide (such as a heterostructure interface \cite{Noda}, shown in the
inset in Fig. 4a), which makes the photon reflected and interacts with the
coupled QD-cavity system again. With precisely adjusting the position of the
element, the final state can be obtained as $\left\vert g\right\rangle
_{1}\left\vert g\right\rangle _{2}\left\vert R\right\rangle $. Therefore, the
QD-QD gate described by $U=e^{\mathrm{{i}}\pi\left\vert g\right\rangle
_{12}\left\langle g\right\vert }$, can be manipulated. Most importantly, this
idea can also be easily extended to realize an \textit{N}-qubit gate with also
only one step. Note that only \textit{N} coupled two-level QD-cavities are
required to realize arbitrary unitary operation on a $2^{N}$-dimensional state
space of \textit{N}-qubits, compared to ($2^{N}-3$) two-qubit controlled gates
without auxiliary qubits \cite{Nielsen}, which is of importance for reducing
the complexity of the physical realization of practical quantum computation
and quantum algorithms.

\begin{figure}[tb]
\centerline{\includegraphics[keepaspectratio=true,width=0.48\textwidth]{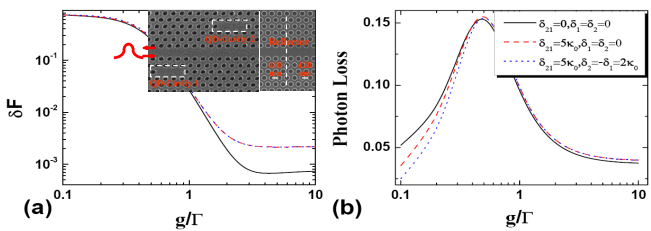}}\caption{(color)
Gate fidelity change ($\delta F\equiv1-F$) (a) and photon loss $P$ (b) of the
two-qubit gate versus $g/\Gamma$. Inset in (a): a configuration of two-qubit
gate operation in which a reflecting element (with $95\%$ reflectivity) is
introduced at the end of the QD-cavity subsystem. Here the carrier frequency
is assumed as $\varpi-2.5\kappa_{0}$ to avoid the EIT-like peaks of two
coupled empty cavities, and a scattering loss of $1\%$ is used in the short
propagation lengths at the slow group velocities. Other parameters are same as
Fig. 2a.}%
\end{figure}

\textit{Gate Fidelity and Photon Loss.}---To exemplify a two-qubit coupled
cavity system, isolated single semiconductor QDs in high-$Q$
small-modal-volume ($V$) photonic crystal nanocavities are potential
candidates, such as self-assembled InAs QDs in GaAs cavities
\cite{Yoshie,Hennessy,Englund,Badolato}, or PbS nanocrystals in silicon
cavities at near $1550$ \textrm{nm} wavelengths \cite{Bose}. For PbS
nanocrystal and silicon nanocavity material system, we use the following
relevant parameters in our calculations: $\gamma_{s}\sim2$ \textrm{MHz},
$\gamma_{p}\sim1$ \textrm{GHz} at cooled temperatures, $Q\sim10^{6}$\ as
reported experimentally \cite{Noda}, $V\sim0.1$ $\mathrm{\mu m}^{\mathrm{3}}$
at $1550$ \textrm{nm}, with resulting single-photon coherent coupling rate $g$
$\sim30$ \textrm{GHz}. To characterize the present gate operation, Figs. 4a
and 4b demonstrate a high two-qubit phase gate fidelity $F$ and a small photon
losses $P$ for different single-photon coupling rates $g$, even under a
non-ideal detuning condition and bad cavity limit. Based on the above
parameters, $F$ can be as high as $0.999$, and $P$ can be as low as $0.04$.
Under various cavity-cavity detunings, both $F$ and $P$ have no significant
degradation. Furthermore, even in the presence of the QD-cavity detuning that
is comparable with the bare cavity linewidth, both $F$ and $P$ keep almost
unchanged. It is remarkable that the on-resonance $P$ is even larger than the
non-resonance case when $g$\ is small. This can be explained by considering
the decay of QDs. When the QDs resonantly interact with cavity modes, the
decay of QDs becomes distinct, which results in an increasing of photon loss.
On the other hand, $P$\ exhibits an increase before a decrease with increasing
$g$, which can be understood by studying the photon loss when the QDs are in
the state of $\left\vert g\right\rangle _{1}\left\vert g\right\rangle _{2}$.
When $g\simeq\Gamma/2$, the absorption strength (resulted $\kappa_{0}$\ from
and $\gamma$) of the input photon by the coupled cavities reaches the maximum.

\textit{Conclusion.}---We have introduced and examined the coherent
interaction of a two-qubit quantum phase gate in a realizable solid-state
nanocavity QED system. The coupling of a nanoscale emitter to the quantized
cavity mode in a coherent array results in unique lineshapes and can give an
important indication of the interactions. Significantly, the fidelity and
photon loss\ of the gate\ are kept within tolerable bounds even in a realistic
semiconductor material system. This provides an approach towards a chip-scale
quantum network for large-scale quantum information storage and computation.

\begin{acknowledgments}
The authors acknowledge funding support from DARPA, and the New York State
Office of Science, Technology and Academic Research.
\end{acknowledgments}

\end{document}